\documentstyle[preprint,prb,aps]{revtex}

\begin{document}
\draft
\title{Relativity of representations in quantum mechanics}
\author{A. C. de la Torre}
\address{Departamento de F\'{\i}sica,
 Universidad Nacional de Mar del Plata\\
 Funes 3350, 7600 Mar del Plata, Argentina\\
dltorre@mdp.edu.ar}
\date{\today}
\maketitle
 \begin{center} Accepted for publication in Am. J. Phys.\end{center}
\begin{abstract}
Only the position representation is used in introductory quantum
mechanics and the momentum representation is not usually presented
until advanced undergraduate courses. To emphasize the relativity
of the representations of the abstract formulation of quantum
mechanics, two examples of representations related to the
operators $\alpha X + (1- \alpha)P$ and $\frac{1}{2}(XP+PX)$ are
presented.
\end{abstract}

\section{INTRODUCTION}
The position representation is adopted in every introductory text
on quantum mechanics. In this representation, the position
observable $X$ for a particle in one dimension is associated with
multiplication by a real variable $x$ and the momentum observable
$P$ with the derivative operator $-i\partial_{x}$. The time
evolution is described by Schr\"{o}dinger's equation whose
solution determines the general state of the system $\psi(x,t)$ or
the stationary states $\psi(x)$ associated with a fixed value of
the energy. (We will use the convention $\hbar=1$ and denote the
derivative operator $d/dx$ by $\partial_{x}$ . We will also
consider only one spatial dimension with a trivial generalization
to three dimensions.)

When students reach an advanced undergraduate quantum mechanics
course, they may arrive with the misconception that the position
representation is the only one or that it is a privileged one.
Then they encounter, as the second choice, the momentum
representation where the observables $(X,P)$ are represented by
$(i\partial_{p},p)$. The students soon learn that this choice is
fully equivalent, not secondary, to the position representation.
Other possible representations are usually ignored. To fully
appreciate the beauty of the mathematical formalism of quantum
mechanics in abstract Hilbert spaces, it is convenient to present
the position and momentum representations as just two particular
representations among an infinite number of choices associated
with all possible observables that can be constructed as functions
of position and momentum. Of course, in practical cases where the
potential depends only on position, the position representation is
more convenient because it leads to simpler differential
equations. And in many cases the momentum representation is more
convenient for stating the initial conditions for the system. The
position and momentum representations are preferred for practical
simplicity, but they are not essential choices of the theory. An
analogous situation occurs when a reference frame, for example,
the center of mass or rest frame, is chosen for simplicity
although any other choice is equally valid.

In this paper we will review how the position and momentum
representations emerge from the abstract formulation of quantum
mechanics, and we will see some examples of other representations
that present some interesting physical and mathematical features.
The representations discussed here can be used to emphasize the
{\em relativity of representations} in the teaching of quantum
mechanics. Many exercises are suggested although not explicitly
stated.

\section{ABSTRACT FORMALISM}
The state of a particle in a one-dimensional space is an element
$\psi$ of an abstract Hilbert space ${\cal H}$ of infinite
dimensions. In addition, the position and momentum observables are
associated with hermitian operators with continuous spectra $X$
and $P$. The physical requirement that the momentum operator be
the generator of translations, that is $X+a{\mathbf
1}=\exp(iaP)X\exp(-iaP)$, leads to the mathematical requirement
that these operators satisfy the commutation relation $[X,P]=i$.
Let $\{\varphi_{x}\}$ and $\{\phi_{p}\}$ denote the two Hilbert
space bases associated with the position and momentum operators,
that is, their eigenvectors correspond to the eigenvalues $x$ and
$p$ respectively. The physical requirement that position and
momentum be independent, in the sense that any momentum is
compatible with any position, requires that these two bases should
be {\em unbiased}, that is, any element $\varphi_{x}$ has an equal
``projection'' along every element $\phi_{p}$. Stated precisely,
the norm of the inner product
$|\langle\varphi_{x},\phi_{p}\rangle|$ should be a constant
independent of $x$ and $p$ and can depend only on the dimension of
the Hilbert space (actually, this constant is undetermined because
the basis elements are not normalizable. This difficulty is
related to the rigorous treatment that will be suggested in
Sec.~\ref{sec:rigged}).

Any state of the system can be expanded with respect to one of the
bases discussed. However, besides the bases associated with the
position and momentum operators, we can define other bases
associated with any observable $F(X,P)$ that depends on position
and momentum and is described by a properly defined hermitian
operator. In the following section we will see how the bases
$\{\varphi_{x}\}$ and $\{\phi_{p}\}$ lead to the position and
momentum representations respectively, and how any other basis can
define a different representation of quantum mechanics.

\section{POSITION, MOMENTUM, AND THE RELATIVITY OF REPRESENTATIONS}
Let us consider the expansion of a state $\psi$ in the basis
$\{\varphi_{x}\}$ associated with the position operator,
\begin{equation}\label{1}
\psi = \! \int_{-\infty}^{\infty}dx\
\langle\varphi_{x},\psi\rangle\varphi_{x}\ .
\end{equation}
Due to the required normalization of $\psi$, the coefficients of
the expansion, given by the function
$\psi(x)=\langle\varphi_{x},\psi\rangle$, must belong to the
Hilbert space ${\cal L}_{2}(\Re)$ of all square integrable complex
functions of a real variable $x$. The function $\psi(x)$ is then
the {\em position representation} of the state. The position
representation results from the isomorphism between ${\cal H}$ and
${\cal L}_{2}(\Re)$, defined by the basis $\{\varphi_{x}\}$. One
can easily determine that in this representation the eigenvectors
of the position and momentum operators are
\begin{eqnarray}
 \varphi_{a}(x)&=&\delta(x-a) \label{2} \\
\phi_{g}(x)&=&\frac{1}{\sqrt{2\pi}}\exp(igx)\ . \label{3r}
\end{eqnarray}
It is important to emphasize to students that in Eqs.~(\ref{2})
and (\ref{3r}), the physically relevant quantities are $a$ and
$g$, whereas $x$ is just a mathematical variable for the functions
in ${\cal L}_{2}(\Re)$.

In an equivalent way we obtain the momentum representation from
the isomorphism between ${\cal H}$ and ${\cal L}_{2}(\Re)$,
defined by the basis $\{\phi_{p}\}$. In this representation, where
the state $\psi(p)=\langle\phi_{p},\psi\rangle$ is an element of
${\cal L}_{2}(\Re)$, the eigenvectors of the position and momentum
operators are given by
\begin{eqnarray}\label{3}
 \varphi_{a}(p)&=&\frac{1}{\sqrt{2\pi}}\exp(-iap) \\
\phi_{g}(p)&=&\delta(p-g)\ .
\end{eqnarray}
Here again, it is important to point out that the physically
relevant quantities are $a$ and $g$, whereas $p$ is just a
mathematical variable.

These two representations arise from two isomorphisms of the
abstract Hilbert space ${\cal H}$, and the isomorphism between
them is defined by the Fourier transformation. This subject is
treated with more or less detail in all advanced books of quantum
mechanics but, in many cases, without reference to the general
abstract Hilbert space. It is however convenient to make this
reference in order to place both representations on an equal
footing and to suggest the existence of many other, equally valid,
possible representations. The relativity of representation implies
some sort of completeness of quantum mechanics in the sense that
it guarantees that the probability distribution for {\em every}
observable $F(X,P)$, represented by a properly defined hermitian
operator, can be obtained from $\psi\in{\cal H}$. To extract this
information, encoded in $\psi$, we must express the state in the
$F$ representation, that is $\psi(f)=\langle
\chi_{f},\psi\rangle$, where $\{\chi_{f}\}$ is the basis
associated with the operator $F(X,P)$.

We will present here two additional representations that turn out
to be interesting from the physical and mathematical point of
view. However, before presenting them, it may be useful to mention
a mathematical difficulty that is often ignored in undergraduate
courses, but that should be presented more rigorously. This
difficulty is sketched in the next section, but can be skipped if
no mathematical rigor is desired.

\section{RIGGED HILBERT SPACE}\label{sec:rigged}
It can be proven that the commutation relation $[X,P]=i$ implies
that the position and momentum operators are unbound and that they
do not have eigenvectors in the Hilbert space. It is a simple
exercise to prove that the assumption of the existence of
eigenvectors of, say $X$, leads to a contradiction when we
calculate the expectation value of the commutator $[X,P]$. Indeed,
the functions given in Eqs.~(2) and (3) or those of Eqs.~(4) and
(5) clearly do not belong to ${\cal L}_{2}(\Re)$ because they are
not square integrable. The bases do not belong to the Hilbert
space ${\cal H}$, but we can anyway expand any element of the
Hilbert space in these bases. In order to achieve this expansion
we must extend the Hilbert space, ${\cal H}\rightarrow{\cal H}'$
to include all such bases. The space so obtained is called a {\em
rigged Hilbert space} or {\em Gelfand triplet} ${\cal
H}^{0}\subseteq{\cal H}\subseteq{\cal H}'$ and is presented in
some advanced texts.\cite{bohm} A rigorous but very clear
exposition of the rigged Hilbert space is given in
Ref.~\onlinecite{bal}.

\section{INTERPOLATING REPRESENTATION}
As an example of another possible representation, we consider the
isomorphism defined by the basis $\{\eta_{\lambda}\}$ of the
eigenvectors corresponding to the eigenvalue $\lambda$ of a family
of operators $S(\alpha)$ that is defined to linearly interpolate
between position and momentum:
\begin{equation} \label{s}
 S(\alpha)=\alpha X+(1-\alpha)P\ , \ \alpha\in[0,1]\ .
\end{equation} In Eq.~(\ref{s}) we have ignored scale factors that
make $X$ and $P$ dimensionless. We have then
\begin{equation}\label{eqn:sn}
 S(\alpha) \eta_{\lambda}=\lambda \eta_{\lambda}\ .
\end{equation}
Using this basis, any state can be expanded as
\begin{equation}
 \psi = \! \int_{-\infty}^{\infty}d\lambda\
 \langle\eta_{\lambda},\psi\rangle\eta_{\lambda}=
 \! \int_{-\infty}^{\infty}d\lambda\
 \psi(\lambda)\ \eta_{\lambda}\ .
\end{equation}
In order to have an expression for $\eta_{\lambda}$ in the
position representation, we must write and solve
Eq.~(\ref{eqn:sn}) in ${\cal L}_{2}(\Re)$. That is,
\begin{equation}
 \left[\alpha
x-i(1-\alpha)\partial_{x}\right]\eta_{\lambda}^{\alpha}(x)
 =\lambda\ \eta_{\lambda}^{\alpha}(x),
\end{equation}
where we have written explicitly the parameter $\alpha$. It is not
difficult to find that the solution of this equation is
$K(\alpha,\lambda) \exp\left[-\frac{i}{2} \frac{\alpha}{1-\alpha}
 \left(x-\lambda/\alpha\right)^{2}\right]$, where the constant
$K(\alpha,\lambda)$ is independent of $x$ but may depend on
$\lambda$ and $\alpha$. We can now choose $K$ such that the
eigenvector $\eta_{\lambda}^{\alpha}(x)$ tends to $\exp(i\lambda
x)$ when $\alpha\rightarrow 0$ and to $\delta(x-\lambda)$ when
$\alpha\rightarrow 1$ as required by Eqs.~(2) and (3). The
appropriate choice for $K$ yields
\begin{equation}
 \eta_{\lambda}^{\alpha}(x)=\frac{1}{\sqrt{2\pi}}
 \frac{\exp\left[i\left(\frac{\lambda^{2}}{2\alpha}+
 \frac{\pi}{4}\right)\right]}
 {\sqrt{1-\alpha}}
 \ \exp\left[-\frac{i}{2}\ \frac{\alpha}{1-\alpha}
 \left(x-\frac{\lambda}{\alpha}\right)^{2}\right]\ .
\end{equation}
Indeed, the limit $\alpha\rightarrow 0$ leads to
\begin{equation}
 \eta_{\lambda}^{0}(x)=
 \frac{\exp\left( i \pi/4\right)}
 {\sqrt{2\pi}} \exp\left(i\lambda x\right)\ .
\end{equation}
For $\alpha\rightarrow 1$, we must use (prove) the unusual
expression for the Dirac delta function
\begin{equation}
 \delta(x)=\lim_{\varepsilon\rightarrow
 0}\frac{1}{\sqrt{\varepsilon}}\
 \frac{\exp \left(i \pi/4\right)}{\sqrt{\pi}}\, \exp
 \left(-i\frac{x^{2}}{\varepsilon}\right)\ ,
\end{equation}
which results in
\begin{equation}
 \eta_{\lambda}^{1}(x)=
 \exp \left(i \frac{\lambda^{2}}{2}\right)\
 \delta\left(x-\lambda\right)\ .
\end{equation}
These eigenfunctions are delta function normalized as is usual for
operators with continuous spectra, that is,
$\langle\eta^{\alpha}_{\lambda},\eta^{\alpha}_{\lambda'}\rangle
=\delta(\lambda-\lambda')$. There are many possible exercises in
this representation. In particular, it is interesting to study the
mathematical transformation between $\psi(x)$ and $\psi(\lambda)$
as function of the continuous parameter $\alpha$ that interpolates
smoothly between the identity and the Fourier transformation.

Instead of the {\em linear} interpolation of Eq.~(\ref{eqn:sn}),
we may consider a phase space {\em rotation} and define the
operator
\begin{equation}
 S(\theta)= X\cos\theta+P\sin\theta\ .
\end{equation}
The treatment for this case is identical to the case just
presented, with the replacement $\alpha\rightarrow\cos\theta$ and
$(1-\alpha)\rightarrow\sin\theta$. One possible interest in this
family of operators follows from the commutation relation
$[S(\theta),S(\theta')]=i\sin(\theta'-\theta)$, indicating that
for $\theta'=\theta+\pi/2$, we have a pair of canonical conjugate
observables that play the same role as position and momentum.

\section{CORRELATION REPRESENTATION}
Another representation of quantum mechanics arises when we build
an isomorphism with the basis $\{\xi_{\gamma}\}$ associated with
the eigenvectors of the {\em correlation operator} defined as the
symmetrized product of position and momentum.
\begin{equation}
 C=\frac{1}{2}\{XP\}=\frac{1}{2}(XP+PX)\ .
\end{equation}
The eigenvalue equation
\begin{equation}\label{eqn:corr}
 C\ \xi_{\gamma}=\gamma\ \xi_{\gamma}
\end{equation}
can be written in the position or momentum representation and
solved to find the associated eigenfunctions. Notice however that
the correlation operator commutes with the parity operator ${\cal
P}$ which changes $X\rightarrow-X$ and $P\rightarrow-P$. Students
can easily prove that this property implies that the eigenvectors
$\{\xi_{\gamma}\}$ must have definite parity, either even
$\{\xi^{g}_{\gamma}\}$ or odd $\{\xi^{u}_{\gamma}\}$ (the upper
index stands for {\em gerade} (even) or {\em ungerade} (odd) under
the parity transformation). The explicit treatment of the above
equation in the position representation provides the two
degenerate solutions.
\begin{eqnarray}\label{eqn:corr2}
 \xi^{g}_{\gamma}(x)&=&K(\gamma)\ |x|^{-\frac{1}{2}+i\gamma}=
 K(\gamma)\ \frac{\exp(i\gamma\ln|x|)}{\sqrt{|x|}} \\
\xi^{u}_{\gamma}(x)&=&K(\gamma)\ {\rm
sign}(x)|x|^{-\frac{1}{2}+i\gamma}=K(\gamma)\ {\rm sign}(x)
 \frac{\exp(i\gamma\ln|x|)}{\sqrt{|x|}}
\ ,
\end{eqnarray}
where $K(\gamma)$ is an arbitrary constant that can be fixed by
requiring the delta function normalization of the eigenvectors.
The momentum representation of the eigenfunctions can be obtained
in the same way, that is, by writing Eq.~(\ref{eqn:corr}) in terms
of $p$ and $\partial_{p}$, or by taking the Fourier transform of
Eqs.~(17) and (18) or, most easily, by noticing that the operator
$C$ in the momentum representation is obtained from the position
representation by replacing $x\rightarrow p$ and taking the
complex conjugate. Therefore, if $\xi_{\gamma}(x)$ is an
eigenfunction in the position representation, then
$\xi^{\ast}_{\gamma}(p)$ is the corresponding eigenfunction in the
momentum representation. These eigenfunctions have the interesting
property that their Fourier transformation is equal to their
complex conjugate.

The correlation operator discussed here has been ignored in most
text books although it is relevant, because it corresponds to an
extra contribution to the uncertainty relations in the improved
version given by Schr\"{o}dinger,\cite{sch}
\begin{equation} \label{ineq}
 \Delta^{2}_{x}\Delta^{2}_{p}\geq\frac{\hbar^{2}}{4}
 +\left(\langle C\rangle-\langle X\rangle\langle
 P\rangle\right)^{2}\ .
\end{equation}
Another interesting property of the correlation operator is that
the term due to the correlation in the inequality (\ref{ineq})
(for general observables) has been related to nonseparability in
compound systems.\cite{tor}

\section{CONCLUSION}
Two possible representations have been sketched in addition to the
position and momentum representations. Many other examples of
representations can be produced and they all illustrate the
importance of the relativity of representations in the abstract
formulation of quantum mechanics. From the mathematical point of
view, this work presents a didactic approach to a general theory
of transformations, because any pair of representations define a
transformation of which, the Fourier transformation is just one
example corresponding to two representations related to two
unbiased bases.

This work received partial support from ``Consejo Nacional de
Investigaciones Cient\'{\i}ficas y T\'ecnicas'' (CONICET),
Argentina.

\end{document}